\documentclass[epj]{svjour}
\usepackage{graphics}

\newcommand \be {\begin{equation}}
\newcommand \ee {\end{equation}}
\newcommand \bea {\begin{eqnarray}}
\newcommand \eea {\end{eqnarray}}

\newcommand \bi {\bibitem}
\newcommand \s {\sigma}

\begin{document}

\title{Aging effects and dynamic scaling in the 3d Edwards-Anderson
spin glasses: a comparison with experiments}

\author{Marco Picco\inst{1}
	\thanks{e-mail: {\tt picco@lpthe.jussieu.fr}} \and
	Federico Ricci-Tersenghi\inst{2}
	\thanks{e-mail: {\tt riccife@ictp.trieste.it}} \and
	Felix Ritort\inst{3}
	\thanks{e-mail: {\tt ritort@ffn.ub.es}}
}

\institute{
	LPTHE, Universit\'e Pierre et Marie Curie, Paris VI,
	Universit\'e Denis Diderot, Paris VII,
	Boite 126, Tour 16, 1$^{\it er}$ \'etage, 4 place Jussieu,
	F-75252 Paris Cedex 05, France
	\and
	Abdus Salam International Center for Theoretical Physics,
	Condensed Matter Group,
	Strada Costiera 11, P.O. Box 586, I-34100 Trieste, Italy
	\and
	Department of Physics, Faculty of Physics,
	University of Barcelona,
	Diagonal 647, 08028 Barcelona, Spain
}
\authorrunning{M. Picco, F. Ricci-Tersenghi and F. Ritort}
\titlerunning{Dynamic scaling in 3d spin glasses}

\date{\today}

\abstract{
We present a detailed study of the scaling behavior of
correlations functions and AC susceptibility relaxations in the aging
regime in three dimensional spin glasses. The agreement between
simulations and experiments is excellent confirming the validity of
the full aging scenario with logarithmic corrections which manifest as
weak sub-aging effects.
}

\maketitle

\section{Introduction}

There is a great interest in the understanding of dynamical effects in
spin glasses. These include magnetization relaxation, aging and
temperature change protocols~\cite{LSNB,HOV,SITGES,NS,RECENT_SG}. The
study of these effects may clarify the nature of spatial effects and
coarsening phenomena in spin glasses, an issue which remains still
poorly understood.

In this paper we present a detailed study of magnetization relaxation
phenomena and aging effects in three dimensional Edwards-Anderson spin
glasses. Our primary goal is to check the validity of the full $t/t_w$
scaling behavior in correlation functions as well as identitying possible
sources of corrections to that behavior by comparing to experimental data.
Dynamical experiments in spin glasses include magnetic relaxation and AC
measurements. Here we will focus our attention on correlation function and
AC susceptibility relaxations. The advantage of studying correlations is
that these are easy to evaluate numerically being also tightly related to
thermoremanent magnetization relaxation experiments. On the other hand, AC
relaxations can be directly compared to experimental results and, to the
best of our knowledge, no results appeared on this point in the literature.

There exist many works on simulations in the
literature~\cite{RIEGER,AMS,KSSR,YOSHINO} studying correlations or
remanent magnetization relaxations and this part of the topic that we
investigate here is certainly not new. What has never been considered
in detail in the past and merits further investigation is the explicit
comparison between simulations and experiments. Having the
experimental results in mind we have tried to apply the same scaling
plots used by the experimentalists to our numerical data. This may
serve as a valuable guide to better understand what properties are
generic to spin glasses and what coarsening scenario accounts for the
collected experimental data.

Magnetic relaxation (or correlation function) and AC experiments give
equivalent information, the advantage of using AC experiments is that
they constitute a very sensitive tool to detect dissipative
processes. When measuring correlations the external timescale $t_w$ is
fixed by the time elapsed after quenching below $T_c$ while in AC
experiments the external timescale is fixed by the inverse of the
frequency of the AC field.  In a full scaling scenario, in the first
class of experiments the relevant scaling variable is $t/t_w$ while in
the second class it is $\omega t$. In what follows we check the
validity of this simple scaling behavior identifying possible sources
of corrections.

\section{The model and the observables}

The Edwards-Anderson model~\cite{EA} is defined by the following
Hamiltonian
\begin{equation}
{\cal H}=-\sum_{(i,j)}\,J_{ij}\s_i\s_j -h\sum_{i=1}^V\s_i \quad ,
\label{eqEA}
\end{equation}
where the indices $i,j$ run from 1 to $V$, the $\s_i$ are Ising spins
($\s_i=\pm 1$) and the pairs $(i,j)$ identify nearest neighbors in a
three dimensional lattice. The exchange couplings $J_{ij}$ are taken
from a random distribution. The simplest choice is a Gaussian
distribution with zero average and finite variance,
\begin{equation}
{\cal P}(J) = \bigl( \frac{1}{2\pi} \bigr)^{\frac{1}{2}} \,
\exp\bigl(-\frac{J^2}{2}\bigr) \quad .
\label{eqP}
\end{equation}

This model displays a spin glass transition at finite temperature
$T_c\simeq 0.95$~\cite{PY,MPR}. A Monte Carlo step corresponds to a
sweep over $V$ randomly chosen spins of the lattice. Monte Carlo
simulations of (\ref{eqEA}) use random updating of the spins with the
Metropolis algorithm. Dynamical experiments use very large lattices
(typical sizes are in the range $L=20-100$) with negligible
finite-size effects for the largest sizes ($L=64$ for magnetization
relaxation experiments and $L=100$ for AC experiments). Correlation
function simulations have been done on a special purpose machine
APE100~\cite{APE} for sizes $64^3$ and averaged over 10 samples. AC
experiments were done for a single sample on a Linux cluster of PC's
for size $L=100$.

Relaxation measurements are done applying a uniform magnetic field and
measuring the decay of the thermoremanent magnetization (hereafter
denoted by TRM), or equivalently, the growth of the zero-field cooled
(ZFC) magnetization. The typical experiment consists in the
following. A sample is fastly quenched below the spin glass transition
temperature for a time $t_w$ ({\it i.e.}\ the waiting time). Then a
uniform small magnetic field $h$ is applied and the growth of the
magnetization measured,
\begin{equation}
\chi_{ZFC}(t_w,t_w+t)=\frac{1}{Vh}\sum_{i=1}^V\s_i(t_w+t) \quad .
\label{eqM}
\end{equation}

Another quantity of interest related to the magnetization which can be 
numerically investigated is the two-time correlation defined by 
\begin{equation}
C(t_w,t_w+t)=\frac{1}{V}\sum_{i=1}^V\s_i(t_w)\s_i(t_w+t) \quad .
\label{eqC}
\end{equation}

The interest of studying correlations instead of zero-field cooled
magnetizations is that they yield the same dynamical
information. Indeed in the stationary regime they are related through
the fluctuation-dissipation theorem (FDT)
\begin{equation}
\chi_{ZFC}(t)=\frac{1-C(t)}{T} \quad .
\label{FDT}
\end{equation}

In AC experiments an oscillating magnetic field
$h(t)=h_0\cos(2\pi\omega t)$ of frequency $\omega=\frac{1}{P}$, where
$P$ is the period, is applied to the system and the magnetization
measured as a function of time
\begin{equation}
M(t)=M_0\cos(2\pi\omega t+\phi) \quad ,
\label{eqMAC}
\end{equation}
where $M_0$ is the intensity of the magnetization and $\phi$ is the
dephasing between the magnetization and the field. The origin of the
dephasing is dissipation in the system which prevents the
magnetization to follow the oscillations of the magnetic field. From
the magnetization one can obtain the in-phase and out-of-phase
susceptibilities defined as
\begin{eqnarray}
\chi'=\frac{M_0\cos(\phi)}{h_0}=\frac{2\int_0^PM(t)\cos(2\pi\omega
t)dt}{h_0} \quad , \label{eqchi1}\\
\chi''=\frac{M_0\sin(\phi)}{h_0}=\frac{2\int_0^PM(t)\sin(2\pi\omega
t)dt}{h_0} \quad . \label{eqchi2}
\end{eqnarray}
The dephasing $\phi$ measures the rate of dissipation in the system
and is given by
\begin{equation}
\tan(\phi)=\frac{\chi''}{\chi'} \quad .
\label{tanfi}
\end{equation}

In numerical simulations the in-phase and out-of-phase
susceptibilities are computed by averaging the right-hand side in
Eqs.(\ref{eqchi1}),(\ref{eqchi2}) over several periods
$P=\frac{1}{\omega}$. This means a very large measurement time for low
frequencies for both experiments and simulations.

In the numerical simulations (both in DC or AC experiments) the
intensity of the probing fields cannot be arbitrarily small because of
the weakness of the signal in comparison to other source of
fluctuations such as finite-size effects (which induce finite-volume
statistical fluctuations for extensive quantities, like the
susceptibility). Consequently, the intensities of the probing magnetic
fields used in numerical simulations are much larger than the
corresponding experimental ones (between 50 and 500 times larger). As
we will comment later, we do not believe that this leads to
conflicting results between simulations and experiments. As soon as
one checks that measurements are done within the linear response
regime then the intensity of the probing field should not be
crucial. Actually the values of the intensity of the fields usually
employed in numerical simulations of TRM experiments are well known to
satisfy linear response~\cite{CKR,PRR_99}.  Note that, in general,
similar difficulties are encountered when analyzing data in both
numerical simulations and real experiments, the main difference is the
absolute magnitude of the time scales one can explore in the two cases
(up to microseconds in simulations and between seconds and days in
experiments).

\section{TRM and correlation function relaxations}

In order to study time scaling in a wide times range (specially in the
aging $t \gg t_w$ regime) experimentalists have measured the decay of
the TRM~\cite{SITGES}, which is strictly related to the
zero-field-cooled one through $M_{\rm ZFC} = M_{\rm FC} - M_{\rm
TRM}$, where the field-cooled magnetization is practically constant in
the glassy phase.

From the numerical simulations point of view the best quantity one can
look at for checking time scaling is the autocorrelation function
$C(t_w,t_w+t)$. It has much less fluctuations than any response to an
external field. Moreover in the quasi-equilibrium regime where the
fluctua\-tion-dissipation theorem (FDT) holds, it gives exact
information on the zero-field-cooled susceptibility via
\begin{equation}
\chi(t_w,t_w+t) = \frac{1-C(t_w,t_w+t)}{T} \quad .
\end{equation}
In the aging regime the connection between correlation and
susceptibility is less trivial. However in the limit of small
perturbing field and large times, where a generalization of the FDT
seems to hold~\cite{FR,MPRR,PRR_99}, a relation between correlation
and susceptibility can still be established. Then, in general,
concerning time scaling, we can safely assume that the autocorrelation
functions decay as the TRM do.

In this section we try to understand which is the best scaling for the
$C(t_w,t_w+t)$ data. The measurements have been taken on 10 samples of a
$64^3$ system, at a temperature $T=0.5$, with waiting times ranging up to
$t_w=10^6$ and measuring times up to $t=10^8$.  We have considered two
different experimental situations, that is with or without an external
magnetic field after time $t_w$ (the evolution up to time $t_w$ being
always with no field), in order to check whether such a small perturbation
may change the dynamical scaling. The external field intensity $h=0.1$ has
been chosen such that the system is in the linear response regime.  If the
magnetic field would be applied during all the experiment we do not expect
any sensible difference with the $h=0$ case.

\begin{figure}
\resizebox{\columnwidth}{!}{\includegraphics{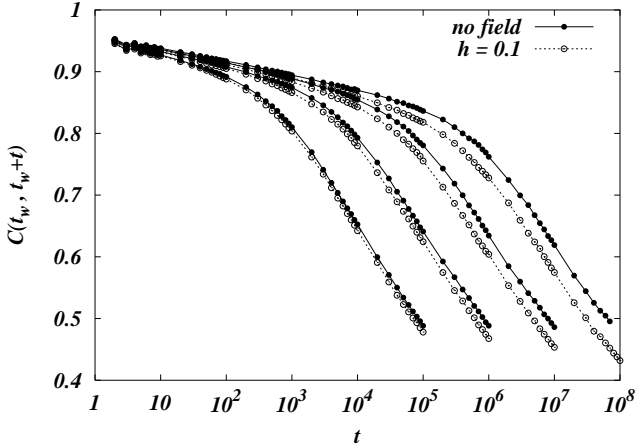}}
\caption{Aging at $T=0.5$ with and without an external magnetic field
after time $t_w$ for a $64^3$ system. Waiting times ($t_w$) range from
$10^3$ (leftmost curves) to $10^6$ (rightmost curves).}
\label{con_senza}
\end{figure}

In Fig.~\ref{con_senza} we show the correlation functions data, with and
without the external magnetic field.  As the time goes on the effect of the
magnetic field seems to accumulate and the differences become larger.  Note
however that, for any given waiting time $t_w$, the correlation curve
presents the two well known regimes~\cite{AMS,RIEGER_REVIEW}: the
quasi-equilibrium one ($t<t_w$) and the aging one ($t>t_w$).

\begin{figure*}
\centering\resizebox{0.6\textwidth}{!}{\includegraphics{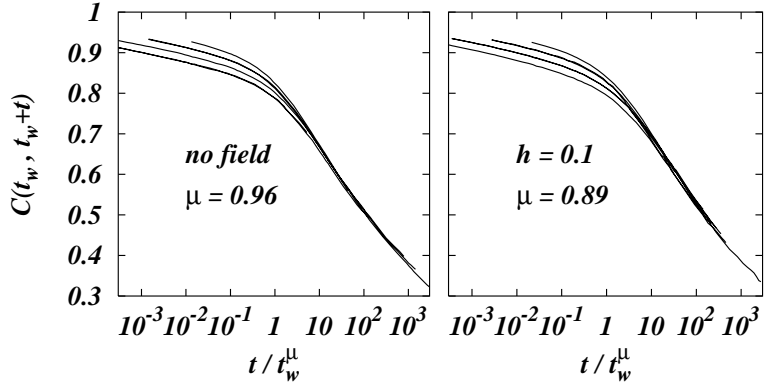}}
\caption{Best scaling in the aging regime ($t>t_w$) for the data
presented in Fig.~\ref{con_senza}}
\label{mu_scal}
\end{figure*}

Because we are mainly interested in the scaling in the aging regime, we
have tried, as the simplest analysis, to collapse the $t > t_w$ data using
$t/t_w^\mu$ as the scaling variable.  The results are shown in
Fig.~\ref{mu_scal} and they clearly show that a value for $\mu$ smaller
than 1 is needed in order to collapse the data in the large times
limit. Note also that the presence of an external field seems to decrease
sensibly the value of $\mu$.  The errors on the estimation of $\mu$ are of
the order of $10^{-2}$.

\begin{figure*}
\centering\resizebox{0.6\textwidth}{!}{\includegraphics{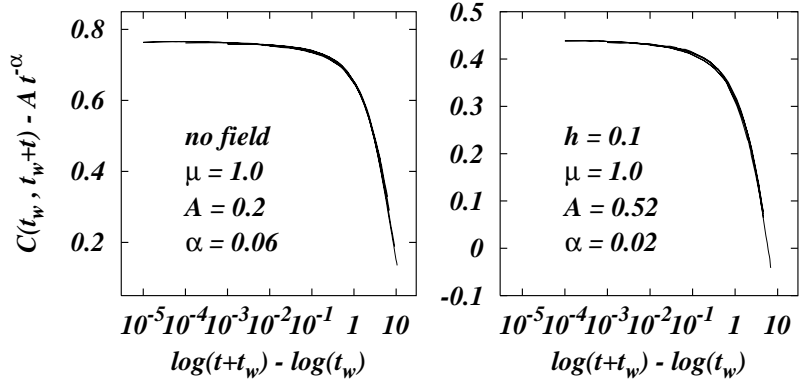}}
\caption{Full aging ($\mu=1$) data collapse for the correlations
functions presented in Fig.~\ref{con_senza}. Note that the use of a
logarithmic scale improves the quality of data collapsing.}
\label{sitges}
\end{figure*}

This numerical result may suggest the presence of a sub-aging regime
in the EA model~\cite{BERTHIER} and it could be interpreted as one
more similarity with real spin glasses, where
$\mu=0.97$~\cite{SITGES}.  However a more careful analysis shows that
the correlation functions are perfectly compatible with a full aging,
that is $t/t_w$, scaling.  In Fig.~\ref{sitges} we present the results
of such an analysis, which has been done following the one performed
on experimental TRM data in reference~\cite{SITGES}.

Few comments are in order.  The best values for the $A$ and $\alpha$
parameters seem to be similar to the experimental ones ($A=0.1$ and
$\alpha=0.02$).  However, because of the lack of a quantitative
criterion for data collapsing, the best collapse is very often
subjective.  In this case we have found that, in order to obtain a
good data collapse, the $\mu$ parameter must be fixed to 1 or very
close to it.  On the other hand, the $A$ and $\alpha$ parameters are
strongly correlated (with a correlation coefficient close to -1) and
they can be changed by a quite large amount without affecting the data
collapse.  Then the errors on these parameters are large. In
Fig.~\ref{sitges} we show the collapse for parameters values being
more or less in the center of the confidence region.  We have also
tried to collapse both sets of data ($h=0$ and $h=0.1$) with the same
parameters, but it was impossible to obtain any reasonable data
collapse.

In Fig.~\ref{sitges} we use the scaling variable $\log(t+t_w)-\log(t_w)$,
even if $t/t_w$ would be the most natural one when full aging holds.  Our
choice is dictated by the need for a comparison with the collapse of
experimental data shown in Fig.3.b of reference~\cite{SITGES}.  There the
scaling variable $[(t+t_w)^{1-\mu}-t_w^{1-\mu}]/(1-\mu)$ is used, which
tends to $\log(t+t_w)-\log(t_w)$ in the $\mu \to 1$ limit.  It is well
known that the goodness of a data collapse may depend on the scales chosen
for presenting the data.  In the present case, in the scaling variable
$\log(t+t_w)-\log(t_w)$ we have better collapses that in the variable
$t/t_w$, because large times are ``compressed''. Note however that both
scaling variables give very good collapses of our data, the same being true
for the experimental data~\cite{VINCENT}.  Anyhow it is worth to note that
the use of the scaling variable $t/t_w$ for checking full aging and
$[(t+t_w)^{1-\mu}-t_w^{1-\mu}]/(1-\mu)$ for checking sub-aging makes the
life harder to the full aging scenario. This without considering the fact
that there could be additional logarithmic corrections to the full $t/t_w$
scaling~\cite{BERTHIER}.

We conclude that it is not easy to obtain precise quantitative
information in order to distinguish the full aging from the sub-aging
scenario. Moreover the presence of an external magnetic field, which
on a first simple analysis seems to change the scaling, is in fact
irrelevant for the scaling and it only changes a little bit the
fitting parameters.

\subsection{A small note on the ZFC susceptibility scaling}

The scaling of the zero-field cooled susceptibility has been already
studied in the past, both directly~\cite{AMS} or via the
fluctuation-dissipation relation which links it to the correlation
functions scaling~\cite{MPRR}.  Here we do not repeat this kind of
analysis.  We simply would like to present some new data regarding the
ZFC susceptibility scaling.

\begin{figure}
\resizebox{\columnwidth}{!}{\includegraphics{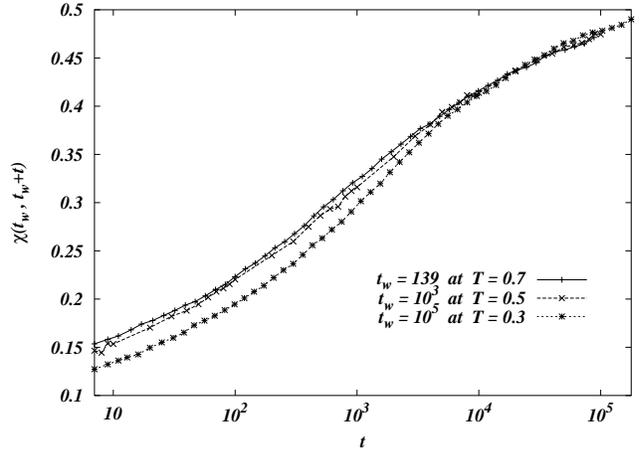}}
\caption{ZFC susceptibilities after a particular experiment (see
text).  If the scaling in Eq.(\ref{EQ_BERN}) would be correct, data in
the figure should collapse.}
\label{FIG_BERN}
\end{figure}

Indeed very recently Bernardi {\it et al.} have proposed the following
scaling for the ZFC susceptibility~\cite{BERNARDI}
\begin{equation}
\chi_{ZFC}(t_w,t_w+t) = \tilde{\chi}(R(t_w),L(t)) \quad ,
\label{EQ_BERN}
\end{equation}
where both length scales grow in an algebraic fashion
\begin{equation}
R(\tau) \propto L(\tau) \propto \tau^{a T} \quad ,
\end{equation}
with an exponent linear in the temperature, like for the
off-equilibrium correlation length~\cite{KSSR,MPRR00,YOSHINO}.  We
found that numerical data obtained with very large simulations are not
compatible with the proposed scaling (see Fig.~\ref{FIG_BERN}).

Our numerical experiments are performed in the following way. For any
given temperature $T_1$ (which takes 3 values in our case
$T_1=0.3,0.5,0.7$), we start the simulation from a random configuration and
we let evolve the system at temperature $T_1$ for a number of MCS $t_w$,
such that $t_w^{T_1}=A$ where $A$ is a constant that we fixed to
$A=10^{1.5}$.  At this time we switch on a small perturbing field, we move
the temperature to $T_2=0.5$ and then we measure the response of the system
(ZFC susceptibility). From this kind of experiment we have obtained many
interesting information on the spin glass low temperature dynamics which
have been published elsewhere~\cite{PRR}. Here we used again this kind of
experiment in order to verify the scaling (\ref{EQ_BERN}) proposed in
Ref.~\cite{BERNARDI}.

By construction we have that in all the three experiments the length
$R(t_w)$ is the same and the $L(t)$ is related to $t$ by the same law,
because after time $t_w$ we always make evolve the system at the same
temperature.  Then if the scaling (\ref{EQ_BERN}) would hold, we should
find a good data collapse in Fig.~\ref{FIG_BERN}, which is not true.  This
behavior can be explained by observing that after the thermalization time
$t_w(T_1)$ the three experimental situations are not identical: they have
developed a similar correlation length, nevertheless the actual
configuration is different and the response to an external perturbation
differs.

In Ref.~\cite{BERNARDI} the authors find a perfect agreement to the
scaling (\ref{EQ_BERN}).  However it must be noted that they use
temperatures in a small range, $T\in[0.5,0.7]$, which corresponds to
the two uppermost curves in Fig.~\ref{FIG_BERN} which indeed almost
coincide.  Violation to the scaling (\ref{EQ_BERN}) can be seen only
at lower temperatures, which where not used in Ref.~\cite{BERNARDI}.

\section{AC susceptibility relaxations}

Let us briefly remind how these experiments are usually done. The
system is quenched to a low temperature (ranging between $0.6$ and
$0.9$ times the value of $T_g$) and the AC susceptibility is
recorded. Typical values of the frequency of the AC field are between
$0.1$ and $1$ Hertz. The amplitude of the probing AC field is of order
$10^{-2}$ to $10^{-1}$ Oersteds deep inside the linear response
regime. The AC susceptibility is then recorded as a function of time
and relaxation is observed on time scales of order $\omega t\sim
1000-5000$ corresponding to several thousands of periods of the AC
field. Measurements are then obtained averaging over several cycles of
the AC field in order to obtain $\chi'$ and $\chi''$ with enough
accuracy (10 cycles is a typical value).

\begin{figure}
\resizebox{\columnwidth}{!}{\includegraphics{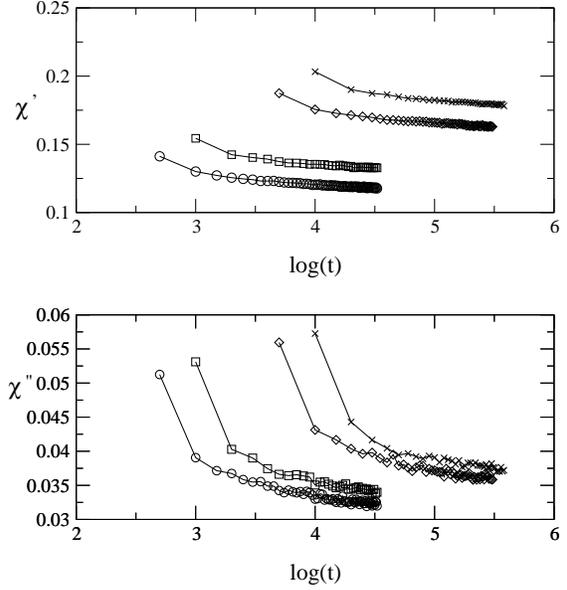}}
\caption{AC susceptibility for $L=100,T=0.6$ and frequencies
$\omega$=$1/P$= 0.02 (circles), 0.01 (squares), 0.002 (diamonds),
0.001 (crosses).}
\label{ACchiT06}
\end{figure}

Having in mind the experimental setup we have done the following
experiment.  Starting from a random configuration we have measured the AC
susceptibility as a function of time for different frequencies of the AC
magnetic field. The frequencies of the field are defined as
$\omega=\frac{1}{P}$ where $P$ is the period of the oscillating field in
Monte Carlo steps.  The results for the {\em in-phase} and {\em
out-of-phase} susceptibilities are shown in Fig.~\ref{ACchiT06} at
temperature $T=0.6\simeq 0.63\,T_c$ and field periods
$P=50,100,500,1000$. Each point in Eqs.(\ref{eqchi1}) and (\ref{eqchi2}) is
obtained by averaging over 10 periods of the field. The curves can be very
well fitted to a power law decays of the type
$\chi(\omega,t)=\chi(\omega,\infty)+At^{-\alpha}$ but the exponents depend
very much on the frequency.  For the largest frequencies the exponent in
both susceptibilities are compatible with a value smaller than $0.1$
although it is difficult to establish its precise value. A similar
difficulty is found in laboratory experiments.

\begin{figure}
\centering\resizebox{\columnwidth}{!}{\includegraphics{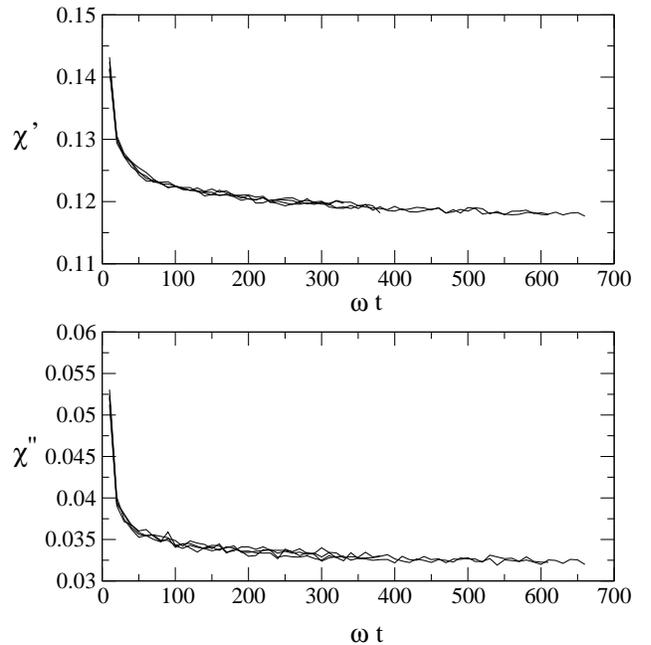}}
\caption{AC susceptibility for $L=100,T=0.6$ and frequencies
$\omega=1/P=0.02,0.01,0.002,0.001$ plotted as a function of $\omega
t$. Following Ref.\cite{SITGES}, susceptibilities have been shifted on
the vertical scale by arbitrary amounts to make them collapse.}
\label{scalACchiT06}
\end{figure}

To make evident the $\omega t$ scaling for the AC experiment we show
in Fig.~\ref{scalACchiT06} the results of Fig.~\ref{ACchiT06} plotted
vs.\ $\omega t$. The values of the susceptibilities in the vertical
scale have been shifted by an arbitrary quantity to make them
coincide. This procedure is exactly the same as done in experiments in
the figure 2 of reference~\cite{SITGES} showing the same qualitatively
results. Note from Fig.~\ref{ACchiT06} that the relaxation of the AC
susceptibility on our time scale is as large as in real experiments,
the relaxing signal being bigger for $\chi''$ than for $\chi'$. This
explains why the study of $\chi''$ is usually preferred in laboratory
experiments. For $\chi''$ the amount of relaxation is nearly equal to
the corresponding stationary ($\omega t\to\infty$) value.

\section{Conclusions}

In this paper we have made a detailed study of the dynamical scaling
behavior of the correlation functions and AC susceptibilities in the
aging regime for spin glasses. The study has been done applying the
scaling behaviors preferred by the experimentalists and comparing them
with the results obtained in numerical simulations of the three
dimensional Edwards-Anderson model.

The general conclusion is that there is a full agreement between the
experimental results measured for TRM decays and simulations for
correlations. This agreement must be understood in the following
sense. All data collected for spin glasses is well compatible with a
full aging scaling scenario with logarithmic corrections. These
corrections are always enhanced in the presence of an external field
and they can be misinterpreted as a subaging scenario.  We are not
aware of any microscopic model displaying aging with full scaling
without logarithmic corrections and it is possible (for not to say
unavoidable) that such corrections are also present in the 3d
Edwards-Anderson model. This fact explains the small subaging effects
measured in both experiments and numerical simulations.

It is also interesting to see how the parameters obtained in the fits
of the decay of the correlation function agree with the equivalent
parameters for the TRM in experiments suggesting a universality in the
dynamical scaling of relaxations which is well captured by the
Edwards-Anderson model.  A similar conclusion is obtained also for AC
experiments which, in general, fulfill a good $\omega t$ scaling
law. The collapse of the AC relaxations on a master curve, by
appropriately shifting them by their stationary values shows a nice
agreement with experimental results and shows how relaxation in
disordered systems are described (at least in a very good
approximation) by a single timescale (corresponding to the waiting
time in TRM experiments or to the inverse of the frequency in AC
experiments).

Still, the big question must be answered. Why this good agreement
between simulations and experiments is not respected when comparing,
at a qualitative level, chao\-tic and memory effects between the
Edwards-Anderson model and real spin glasses ?~\cite{KYT,PRR,YLB}. It is
plausible that for time scales short compared to experimental time
scales some dynamical effects (in particular, chaos and memory) are
not seen in simulations while the full $t/t_w$ scaling (with
logarithmic corrections) is already present in this regime. Further
theoretical, numerical and experimental studies are needed to clarify
this controversial issue.

{\bf Acknowledgements} We are grateful to E. Vincent for useful
discussions.  M.P. and F.R. acknowledge financial support from a
French-Spanish collaboration (Picasso program and Acciones Integradas
Ref. HF1998-0097).

\end{document}